\documentclass[10pt, letterpaper]{article}

\usepackage[a4paper, total={6.5in, 8in}]{geometry}
\usepackage{graphicx}
\usepackage{xcolor}
\usepackage{float}
\begin{document}


\noindent\fontsize{16}{24}\selectfont\textbf{Data-driven brain network models predict individual variability in behavior}
\\

\fontsize{14}{22}\selectfont\noindent
Kanika Bansal\textsuperscript{1,2,3},
John D. Medaglia\textsuperscript{4,5},
Danielle S. Bassett\textsuperscript{6,7,5,8},
\newline Jean M. Vettel\textsuperscript{2,6,9},
Sarah F. Muldoon\textsuperscript{1,10,*}
\\

\fontsize{10}{14}\selectfont{\noindent$^1$ Department of Mathematics, University at Buffalo, SUNY, Buffalo, NY, USA 
\\
$^2$ U.S. Army Research Laboratory, Aberdeen Proving Ground, MD, USA
\\
$^3$ Department of Biomedical Engineering, Columbia University, New York, NY, USA
\\
$^4$ Department of Psychology, Drexel University, Philadelphia, PA, USA
\\
$^5$ Department of Neurology, Perelman School of Medicine, University of Pennsylvania, Philadelphia, PA, USA
\\
$^6$ Department of Biomedical Engineering, University of Pennsylvania, Philadelphia, PA, USA 
\\
$^7$ Department of Electrical and Systems Engineering, University of Pennsylvania, Philadelphia, PA, USA
\\
$^8$ Department of Physics and Astronomy, University of Pennsylvania, Philadelphia, PA, USA
\\
$^9$ Department of Psychological and Brain Sciences, University of California, Santa Barbara, CA, USA
\\
$^{10}$ Computational and Data-Enabled Science and Engineering Program, University at Buffalo -- SUNY, Buffalo, NY, USA
\\

\noindent*Corresponding author e-mail: smuldoon@buffalo.edu}
\newpage
\section*{Abstract}

The relationship between brain structure and function has been probed using a variety of approaches, but how the underlying structural connectivity of the human brain drives behavior is far from understood.  To investigate the effect of anatomical brain organization on human task performance, we use a data-driven computational modeling approach and explore the functional effects of naturally occurring structural differences in brain networks. We construct personalized brain network models by combining anatomical connectivity estimated from diffusion spectrum imaging of individual subjects with a nonlinear model of brain dynamics.  By performing computational experiments in which we measure the excitability of the global brain network and spread of synchronization following a targeted computational stimulation, we quantify how individual variation in the underlying connectivity impacts both local and global brain dynamics. We further relate the computational results to individual variability in the subjects' performance of three language-demanding tasks both before and after transcranial magnetic stimulation to the left-inferior frontal gyrus.  Our results show that task performance correlates with either local or global measures of functional activity, depending on the complexity of the task. By emphasizing differences in the underlying structural connectivity, our model serves as a powerful tool to predict individual differences in task performances, to dissociate the effect of targeted stimulation in tasks that differ in cognitive complexity, and to pave the way for the development of personalized therapeutics.

\section*{Author summary}

How the organization of the brain drives human behavior is an important but open question in neuroscience. Recent advances in non-invasive imaging techniques and analytical tools allow one to build personalized brain network models that simulate an individual's brain activity based on the underlying anatomical connectivity.  Here, we use these computational models to perform virtual experiments in order to predict individual performance of three different language-demanding tasks.  For each individual, we build a subject-specific brain network model and measure the global brain activation and patterns of synchronized activity after a targeted computational stimulation.  We find that depending on the complexity and type of the language task performed, either global or local measures of brain dynamics are able to predict individual performance.  By emphasizing individual differences in human brain structure, the model serves as a powerful tool to predict cognitive task performance and to promote the development of personalized therapeutics.

\newpage

\section*{Introduction}
Cognitive responses and human behavior have been hypothesized to be the outcome of complex interactions between regional populations of neurons \cite{{Mcintosh1999},{Misic2016}} and show significant variability across individuals.  While certain patterns of brain activity are robust \cite{Bressler2010}, many patterns change with learning and aging \cite{{Zatorre2012},{Bassett2011b},{Zacks2006},{Nestor2009},{TELESFORD2016}}, and an underlying inter-subject variability in neural activity has been observed \cite{{Mueller2013},{Schmaelzle2017},{Telesford2017},{GARCIA2017},{Nestor2013}}.  Importantly, the underlying anatomical connectivity of the brain provides a crucial backbone that drives neuronal dynamics and thus behavior \cite{{Zatorre2012},{Roberts2013},{Reijmer2013}}. Given recent and ongoing advancements in imaging techniques, such as diffusion spectrum imaging (DSI), which estimates the presence of white matter tracts connecting brain regions, mesoscale maps of anatomical brain connectivity can now be obtained \cite{{Roberts2013},{Vettel2017}}. While differences in brain connectivity have long been known to exist between diseased and healthy populations \cite{{Bassett2009},{vandenHeuvel2013},{Gong2015}}, recent findings indicate measurable differences in patterns of white matter connectivity across healthy individuals \cite{{BASSETT2011},{Yeh2016},{Kahn2017},{Powell2017}}. Although work is beginning to link individual variability in white matter structure, functional activity, and task performance \cite{{Medaglia2017},{KELLER2016},{vandenBos2015},{Mckenna2015},{Brown2015},{Muraskin2016},{Muraskin2017}}, there is currently no standard methodology for evaluating the interplay between the brain's structural topology, dynamics, and function, and many open questions remain about how these features are coupled.

The new field of network neuroscience \cite{Bassett2017} provides a coherent framework in which to model and investigate this coupling. In the context of modeling human brain networks, network nodes can be chosen to represent brain regions and network edges can represent either physical connections (anatomical networks) or statistical relationships between nodal dynamics (functional networks) \cite{{Bassett2009},{FELDT2011225},{Bassett2017}}. Analytical tools from network science have been successfully utilized to quantify both structural and functional brain network organization and to gain insight into topics as diverse as brain development \cite{CAO201476}, disease states \cite{{vandenHeuvel2013},{Gong2015},{MENON2011483}}, learning \cite{Zatorre2012}, and intelligence \cite{Cole2012}. 

Combining a network representation of the brain with computational modeling of brain dynamics allows one to further investigate the links between brain structure, dynamics, and performance by providing a controlled environment in which to perform \emph{in silico} experiments and make predictions about real-world brain function. Using experimentally obtained structural brain network data combined with biologically motivated computational models of brain dynamics \cite{Breakspear2017}, one can build personalized brain network models of human brain activity \cite{{Honey2009},{SANZLEON2015},{Bansal:2018ug}}. This modeling approach has been used to gain insight into structure-function relationships in disease populations \cite{{Stefanovski2016},{Adhikari2015}}, perform virtual lesioning or resection experiments \cite{{Alstott2009},{Hutchings2015},{Sinha2017}}, and assess the differential impact of stimulation to different brain regions \cite{{Muldoon2016},{Spiegler2016}}. 

Here, we use this data-driven modeling approach to investigate the interplay between structural variation, brain activity, and task performance. Our computational model is built upon subject-specific connectomes that are combined with biologically informed Wilson-Cowan nonlinear oscillators (WCOs) \cite{{Wilson1972},{Wilson1973}} to produce simulated patterns of personalized brain activity. Additionally, this controlled computational environment allows us to perform targeted stimulation experiments -- motivated by actual laboratory experiments -- and quantify the emerging neural activity patterns that represent both global brain network activation and task-specific local subnetwork activation. Using this model, we construct computational measures in order to relate structure and individual performance variability across a cohort of ten healthy individuals who performed three language-demanding cognitive tasks before and after transcranial magnetic stimulation (TMS) \cite{Ferreri2013} targeted at the left inferior frontal gyrus (L-IFG) of the brain. The performed tasks involved verb generation, sentence completion, and number reading, and are known to vary in their cognitive complexity \cite{{KRIEGERREDWOOD201424},{Snyder2008}}.

Based upon the patterns of simulated brain activity driven by the structural networks in our model, we find that task performance can be explained and predicted by either local or global circuitry depending on the complexity of the task. Further, we observe that, post experimentally applied TMS, the correlation between model output and task performance is weakened. Finally, we show that the eigenspectrum of the observed structural brain networks plays a key role in global brain dynamics which can additionally provide predictive insight into performance of some, but not all, tasks. Taken together, our results reveal that using personalized brain network models to simulate brain dynamics provides an important tool for understanding and predicting human performance in cognitively demanding tasks and represents an important step towards the development of personalized therapeutics.
 
\section*{Results}
\begin{figure} 
\centering
\includegraphics[width=0.8\linewidth, keepaspectratio]{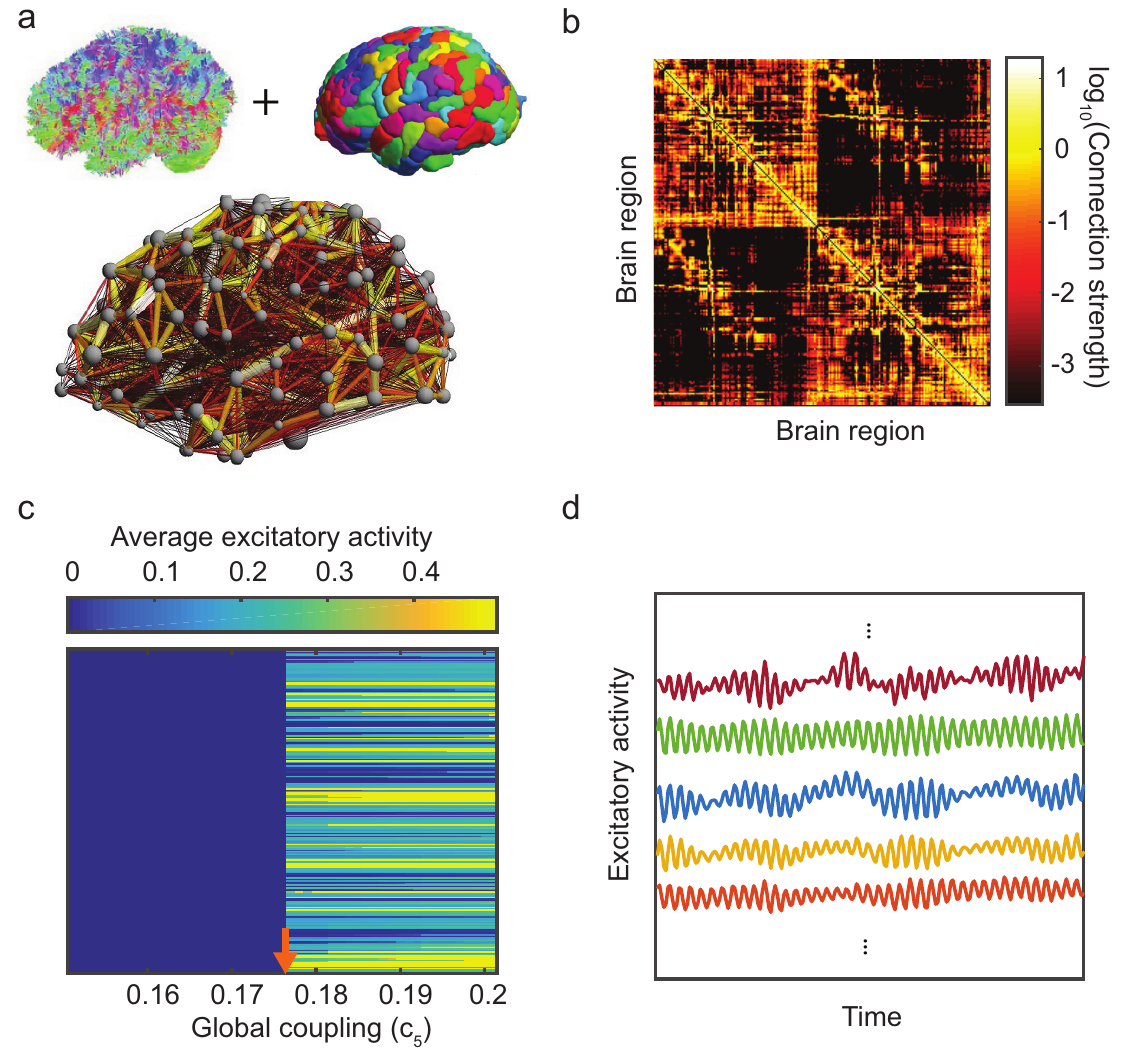}
\caption{{\bf Data-driven brain network models.} 
(a) The brain connectivity used as a basis for the computational model is obtained by combining tractography estimates from diffusion spectrum imaging data of a specific individual's brain and a parcellation of the brain into 234 regions. (b) The resulting anatomical connectivity matrix where entries indicate the density of connections between two brain regions.  (c) The dynamic state of brain activity can be tuned by the global coupling parameter, $c_5$. As this parameter crosses a threshold, a sudden transition to an excited state is observed as marked by the red arrow. (d) Representative excitatory dynamics of selected brain regions in the excited state, demonstrating a rich spectrum of temporal activity that is driven by the underlying anatomical connectivity.}
\label{fig:schematic}
\end{figure}

In order to assess the link between variability in brain connectivity, activity patterns, and behavior, we build data-driven personalized brain network models of ten individuals who performed three language-demanding cognitive tasks before and after TMS targeted at the L-IFG (see Materials and Methods). Subject-specific anatomical connectivity was derived from  DSI imaging data, combined with a parcellation of the brain into 234 regions (network nodes) based on the Lausanne atlas \cite{Hagmann2008} (Fig. \ref{fig:schematic}a). As a result, we obtained a weighted connectivity matrix whose entries represent the strength of the connection between two brain regions (Fig. \ref{fig:schematic}b). The dynamics of each brain region are then modeled using nonlinear Wilson-Cowan oscillators (WCOs), coupled through a subject-specific anatomical brain network (Materials and Methods).  A WCO is a biologically motivated model of local brain activity, developed to describe the mean behavior of small neuronal populations \cite{Wilson1972}. The model therefore simulates a specific individual's spatiotemporal macro-level brain activity.

The only model feature that varies across subjects is the underlying anatomical connectivity matrix, $A$, providing a causal link between differences in regional brain dynamics and structural organization. The global strength of the coupling between different brain regions via $A$ is controlled by a global coupling parameter ($c_5$; Materials and Methods).  The dynamical state of  the brain can be tuned by varying this parameter as shown in Fig. \ref{fig:schematic}c which depicts the average excitatory dynamics as a function of global coupling.  When the global coupling parameter exceeds a threshold value ($c_5^{T}$), the brain dynamics abruptly transition to an active state (Fig. \ref{fig:schematic}d). Previous work has shown that the computational model is particularly sensitive to the point at which model dynamics undergo this transition: the value of $c_5^{T}$ at which the transition takes places is subject-specific, and the inter-subject variability in $c_5^{T}$ is greater than that seen using anatomical networks derived from different scans of the same individual  \cite{Muldoon2016}. The transition value $c_5^{T}$ quantifies the ease (or difficulty) with which a brain network can be excited, and we use this as a parameter to measure differences in global network dynamics between individuals.
 
\subsection*{Individual variability: model and experiments}
\begin{figure} 
\centering
\includegraphics[width=0.8\linewidth, keepaspectratio]{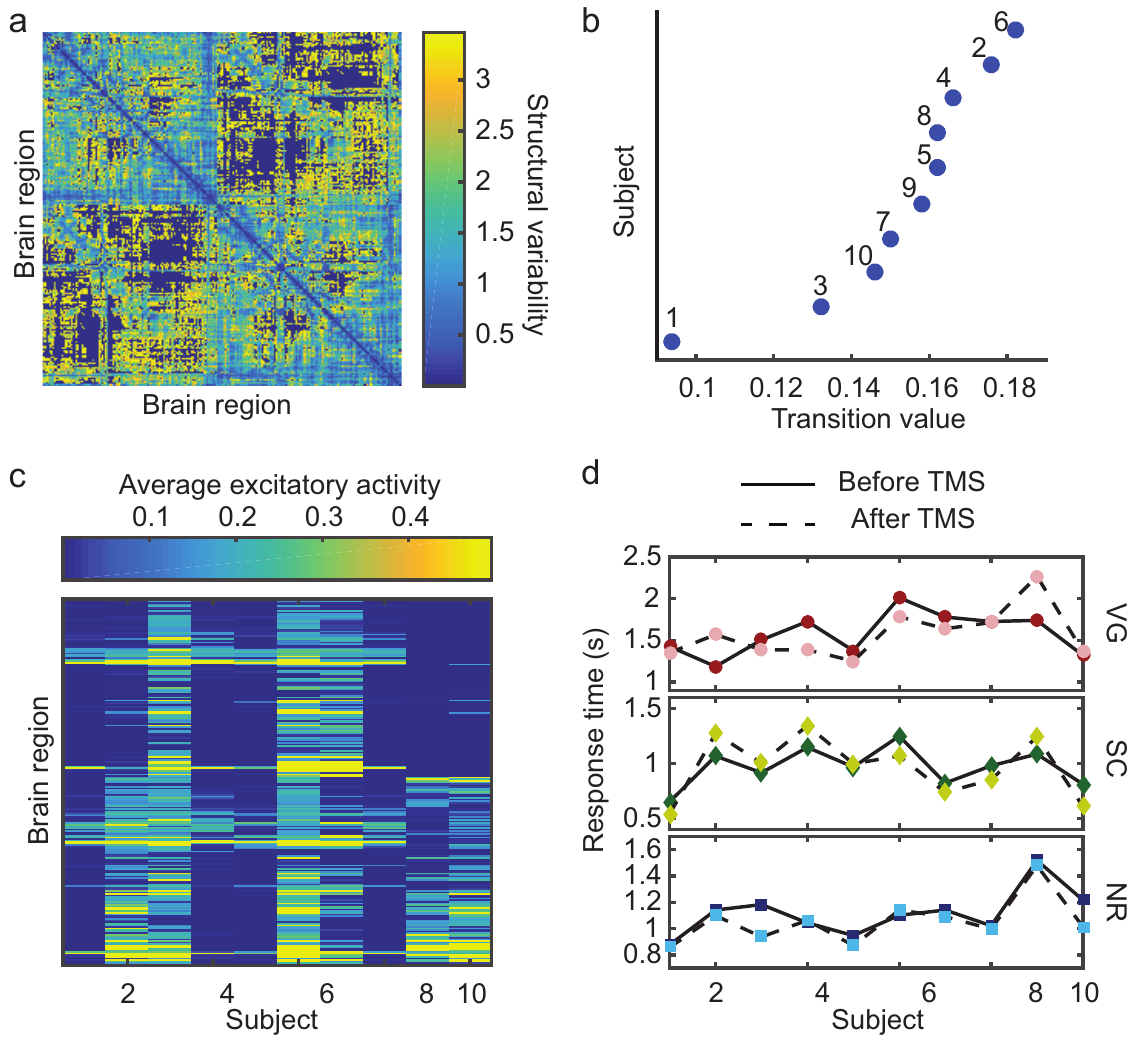}
\caption{{\bf Assessing variability in structure, dynamics, and performance.} 
(a) Variability in the connection strengths between pairs of brain regions across ten subjects, measured as the standard deviation of edge weights between brain regions across individuals, normalized by the mean of edge weights. Although there are few regions of low variability (blue), there are a significant number of regions with moderate to high variability (green to yellow). (b) The global coupling transition value ($c_5^{T}$) varies across our cohort of ten individuals. (c) Spatial patterns of excitatory activity vary across individuals due to regional variations in the structural connectivity between subjects. Each column represents the temporal average of the excitatory activity across brain regions for a given individual in the excited state and is unique in terms of the overall activity pattern. (d) Cognitive task performance for the ten subjects was assessed by experimentally measuring the response times for three language-demanding tasks: verb generation (VG), sentence completion (SC), and number reading (NR), both before (solid lines) and after (dotted lines) a targeted transcranial magnetic stimulation (TMS) to the left-inferior frontal gyrus (L-IFG) of the brain.}
\label{fig:variability}
\end{figure}

Because we are interested in linking variability in brain structure and behavior, we first measured the extent of individual variability in anatomical connectivity, simulated brain activity, and task performance in our data set. Across our cohort of subjects, we observed measurable variability in anatomical network structure as seen in Fig. \ref{fig:variability}a, which shows the standard deviation in edge weights between two brain regions, normalized by the mean edge weight. This structural variability is also manifested in the simulated brain activity depicted in Fig. \ref{fig:variability}b-c.  We observe variability across individuals in the global coupling transition values ($c_5^{T}$) (Fig. \ref{fig:variability}b) as well as in the specific patterns of brain activity at this transition (Fig. \ref{fig:variability}c). Since the nonlinear WCOs are all identical in the model, these observed differences in simulated brain activity are a direct result of variation in the underlying anatomical connectivity. 

Finally, we assessed the extent of variability in the cognitive performances of individuals across three language-demanding tasks: (i) verb-generation (VG); (ii) sentence-completion (SC); and (iii) number-reading (NR). Performance was measured as the median response time across $50$ trials (see Materials and Methods) and shows variability across both subjects and tasks (Fig. \ref{fig:variability}d). Moreover, performances were altered after transcranial magnetic stimulation (TMS) was applied to the left inferior frontal gyrus (L-IFG). The L-IFG, also known as Broca's area, is traditionally believed to play an important role in language comprehension and syntactic processing, and specifically in the selection and retrieval of words \cite{Dronkers2004}. It is therefore expected that external stimulation to this region should affect task performance. However, it is important to note that while subject performance on a task did often change after TMS, we did not consistently observe an improvement (or degradation) in task performance across individuals, nor was the effect consistent between tasks within a given subject. This suggests that although the L-IFG plays an important role in the context of language comprehension, the actual cognitive response reflects contributions from a larger part of the brain network.

Given the observed variability in the structural, dynamical, and behavioral aspects of our data, we next focused on assessing how this variability was related across these three domains. We therefore examined network features measured at both the global network level (using the entire brain network) and within task-specfic subnetworks that were selected to represent the specific circuitry involved in task completion and asked how these measures related to task performance.

\subsection*{Global network features predict certain task performances}
\begin{figure}
\centering
\includegraphics[width=0.85\linewidth] {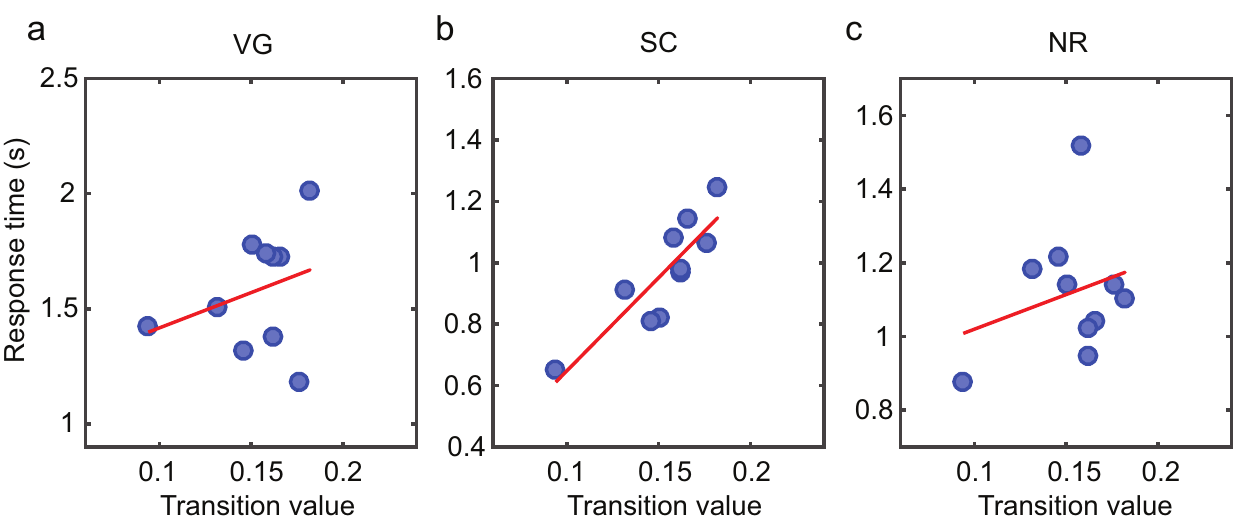}
\caption{{\bf Correlation between model transition threshold and task performance.} 
Task performance versus model transition values, $c_5^{T}$, for three tasks: (a) verb generation (VG), (b) sentence completion (SC), and (c) number reading (NR). The red lines represent a linear fit to the data for visual guidance. The corresponding Pearson correlation coefficients are given in Table \ref{table:corr_pre_tms}. There is a significant correlation between the global transition value and task performance for the SC task ($r=0.86$, $p=0.001$).}
\label{fig:pre_tms1}
\end{figure}

We first examined the relationship between global network properties and task performance by estimating the correlation between global brain activation and task performance. For each subject, we measured the threshold value of the global coupling parameter ($c_5^{T}$) at which the individual's brain transitions to the excited state, and we calculated the correlation between this value and their performance on each task (before the application of TMS).  As seen in Fig. \ref{fig:pre_tms1} and Table \ref{table:corr_pre_tms}, we observed a significant positive correlation ($r = 0.86$, $p = 0.001$) between model transition values and task performance in the sentence completion (SC) task. Thus for the SC task, individuals with a lower value of $c_5^{T}$ (more easily excitable brain) are likely to perform better (as measured through a short response time).  However, we did not observe a significant correlation in the verb generation (VG) or number reading (NR) tasks, indicating that the performance of these tasks cannot be predicted by a global network property. To ensure that these results were dependent upon the organization of the subject-specific anatomical connectivity used as a basis for the computational model, for each individual, we created randomized brain networks by preserving the distribution of edge weights but randomly reassigning connection strengths between brain regions (Materials and Methods).  We recalculated the $c_5^{T}$ values for simulations using these randomized connectivity matrices, but did not observe any significant correlations between transition values and task performance in this case.

\begin{table}[!ht]
\centering
\caption{
{\bf Correlations between model features and task performance.}}
\begin{tabular}{|l|cc|cc|cc|}
\hline
&\multicolumn{2}{c|}{VG}  &  \multicolumn{2}{c|}{SC} &  \multicolumn{2}{c|}{NR}\\ 
 Model feature& r & p & r & p& r & p  \\ \hline
Transition value& 0.30& 0.39& \textbf{0.86} &  \textbf{0.001} & 0.27 & 0.45\\ \hline
Functional effect (global brain) & -0.04& 0.91&0.39 & 0.26 & 0.23 &  0.52\\ \hline
Functional effect (task circuit) & 0.42& 0.22 & \textbf{0.73} & \textbf{0.017} & \textbf{0.74} &  \textbf{0.016}\\ \hline
Functional effect (outside the task circuit) &-0.06& 0.87 & 0.38 &0.29 & 0.18 & 0.62\\ \hline
\end{tabular}
\begin{flushleft} The variables $r$ and $p$ denote the Pearson correlation coefficient and associated \textit{p}-value, respectively. VG = verb generation, SC = sentence completion, and NR = number reading. 
\end{flushleft}
\label{table:corr_pre_tms}
\end{table}

\begin{figure}
\centering
\includegraphics[width=.7\linewidth, keepaspectratio]{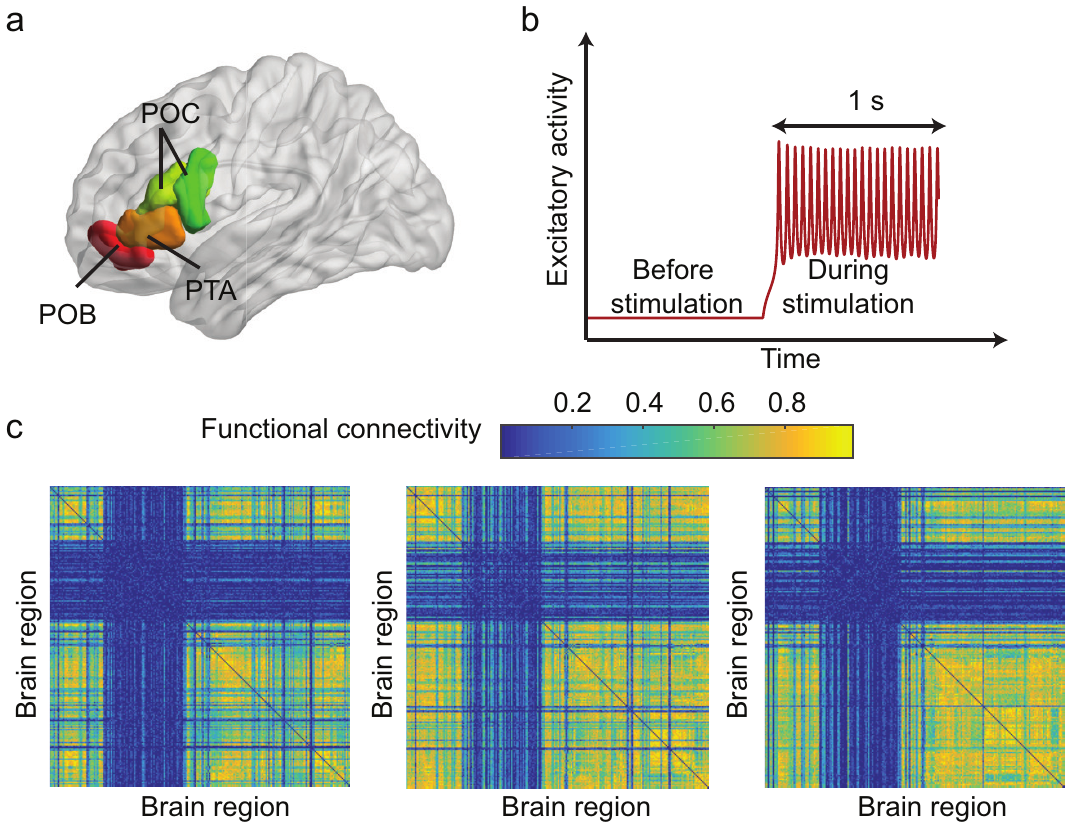}
\caption{{\bf Effects of targeted stimulation.} 
(a) The left-inferior frontal gyrus (L-IFG) is composed of four brain regions in the parcellation scheme used in this study. These regions are \textit{pars orbitalis} (POB), \textit{pars triangularis} (PTA) and \textit{pars opercularis} (POC, two regions). (b) Effect of computational stimulation to a single brain region. The stimulation pushes the dynamics of the region into a limit cycle (oscillations). (c) Functional connectivity, calculated as the pairwise maximum normalized correlation between brain regions for three subjects as a result of stimulating the L-IFG. The observed connectivity patterns were highly variable across subjects.}
\label{fig:functional}
\end{figure}

To further explore the link between global brain dynamics and behavior, we additionally assessed the relationship between specific patterns of brain activity and task performance.  Since the L-IFG is involved in controlled language processing \cite{{Whitney2011},{Costafreda2006},{Hannah2007}}, one can argue that the pattern of brain synchronization as a result of targeted stimulation to the region might also be predictive of task performance. We therefore computationally stimulated the brain regions comprising the L-IFG (Fig. \ref{fig:functional}a), and quantified how the stimulation spread throughout the global brain network (see Materials and Methods for details).  As shown in Fig. \ref{fig:functional}b, computational stimulation pushes the dynamics of the region into oscillatory activity which then drives the functional dynamics of other brain regions through the underlying structural connections. We measure the resulting pattern of brain activity by calculating the pairwise \textit{functional connectivity} using the maximum normalized correlation between brain regions \cite{{Muldoon2016},{Kramer2009},{Feldt2007}} (Materials and Methods).  Due to the variability in the subject-specific structural connectivity matrices, the resulting patterns of functional connectivity differ between individuals.  Fig. \ref{fig:functional}c shows the functional connectivity for three subjects as a result of L-IFG stimulation. The change in the patterns of how the stimulation spreads through the network is evident. We measure the extent of the global spread of synchronization by calculating the functional effect \cite{Muldoon2016} which measures the average value of synchronization across the entire brain (Materials and Methods).  Interestingly, unlike our observation with the global coupling parameter, the global functional effect does not show a significant correlation with task performance for any of the cognitive tasks (Table \ref{table:corr_pre_tms}).

\subsection*{Localized task-specific circuits predict other task performance}
While we did observe a significant correlation between the global threshold value and task performance for the SC task, we saw no correlations between global brain dynamics and task performance in the remaining two tasks, and the global functional effect was not correlated with performance in any of the tasks.  However, the three language tasks performed in this study differ in semantic demands, and the absence of a significant correlation for the VG and NR tasks could be due to either a drastically different cognitive mechanism for performing these tasks, or the dependence of these tasks on a more localized brain circuit.  To investigate the latter possibility, we investigated the role of task-specific subnetworks in task performance.

\begin{figure}
\centering
\includegraphics[width=0.85\linewidth]{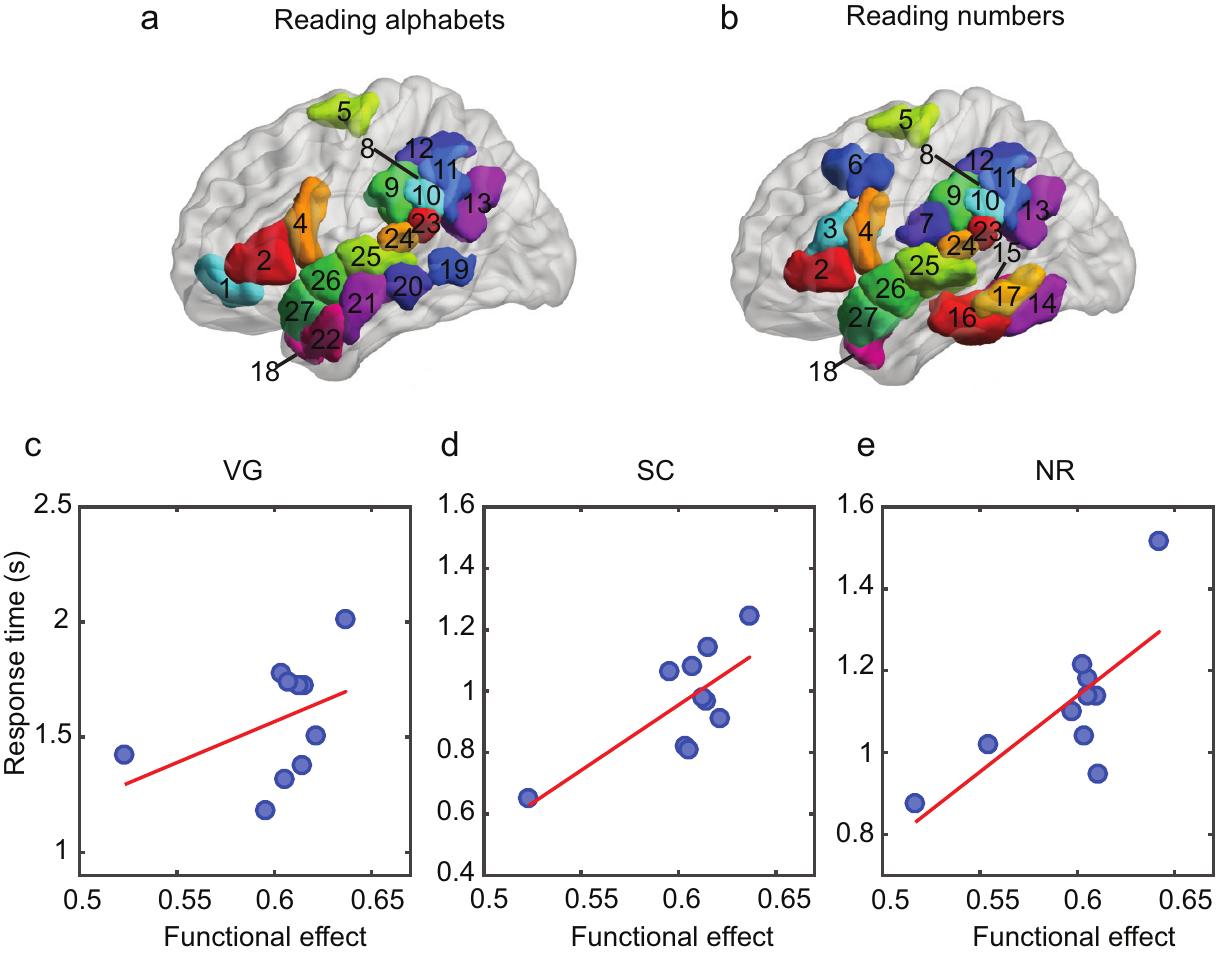}
\caption{{\bf Task-specific circuits.} 
(a) Alphabet reading task circuit used in the analysis of VG and SC tasks. (b) Number reading task circuit used in the analysis of the NR task. Brain regions in the circuits are 1: \textit{pars orbitalis}, 2:  \textit{pars triangularis}, 3:  \textit{pars opercularis-1}, 4:  \textit{pars opercularis-2}, 5:  \textit{superiofrontal-9}, 6:  \textit{caudal middle frontal-1}, 7:  \textit{postcentral-7}, 8-12:  \textit{supramarginal}, 13:  \textit{inferioperietal-3}, 14-15:  \textit{fusiform-2 and 3}, 16-17:  \textit{inferio temporal-2 and 3}, 18:  \textit{temporal pole}, 19-22:  \textit{middle temporal}, 23-27:  \textit{superior temporal}. (c)-(e)Task performance versus the functional effect within task-specific cirucuits for three tasks: (a) verb generation (VG), (b) sentence completion (SC), and (c) number reading (NR). The red lines represent a linear fit to the data for visual guidance. The corresponding Pearson correlation coefficients are given in Table \ref{table:corr_pre_tms}. There is a significant correlation between the functional effect and task performance for the NR task ($r=0.74$, $p=0.016$).}
\label{fig:circuits}
\end{figure}

To construct a task dependent, spatially localized measure, we follow the work of Roux et al. \cite{Roux2008} to identify the possible brain regions involved in reading alphabets and numbers.  Roux et al. performed experimental electrostimulation to spatially map the brain regions that were differentially involved in reading both alphabets and numbers.  We mapped these regions to the Lausanne atlas and constructed two task circuits: one involved in VG and SC (alphabets-related, Fig. \ref{fig:circuits}a), and one involved in NR (number-related, Fig. \ref{fig:circuits}b). These circuits are also consistent with other studies mapping brain regions involved in language processing \cite{{Dronkers2004},{Turken2011},{NathanielJames2002},{Holland2001}}.  Note that both of these subnetworks are contained entirely in the left hemispheric language network.

We then calculated the functional effect within these task circuits (averaging the functional connectivity values only within the subnetwork as opposed to the entire brain network as done previously) and correlated this local measure with task performances (Fig. \ref{fig:circuits}c-e and Table \ref{table:corr_pre_tms}). We observed a significant positive correlation between the task-specific functional effect and task performance for the NR task ($r=0.74$, $p=0.016$), indicating that performance on the NR task depends on the localized spread of activation throughout the task sub-network and is less dependent on the global brain network structure. Individuals with a lower functional effect (less synchronization within the task circuit) also have a lower response time (better performance).  When synchronization within the task circuit is increased (a high functional effect), task performance degrades, suggesting that high levels of synchronized activity within the task circuit could potentially impede the ability of the circuit to perform localized computations necessary for task completion.

If we compare the functional effect measured only within brain regions outside of the task circuit, the significant correlation with the NR task is lost (Table \ref{table:corr_pre_tms}), indicating the specificity of the task circuit in the model.  Although we also observed a significant correlation between the task-specific functional effect and task performance for the SC task, this result is driven by a single subject and does not hold if this subject is removed from the data set.  Performing the analysis on a larger data set would therefore be necessary to confirm this finding for the SC task. No significant correlations were observed for the VG task.

To validate the specificity of our selected task circuits, we constructed 10,000 random sub-networks by randomly selecting the same number of brain regions as in each task circuit and then calculated the functional effect within these random circuits after stimulating the L-IFG (Materials and Methods). The randomized circuits had a variable degree of overlap with the actual task circuit (0 to 35\%). We observed that only 1.8\% of the randomly selected circuits gave significant correlations ($r>0.5$ and $p<0.05$) between the circuit-specific functional effect and task performance, which we estimate to be the false positive rate in our computational predictions. This low error rate signifies that the observed significant correlation in the NR task is due to the selection of brain regions in the task-specific circuit.

We also verified that the observed effects were related to our choice of stimulating the L-IFG as opposed to some other brain structure.  We chose different sets of brain regions (equal in size to the number of brain regions that compose the L-IFG) that were randomly distributed within the task circuit and applied targeted computational stimulation to these randomly selected regions.  When randomly selected brain regions were stimulated, we did not observe a significant correlation between the task performance and functional effect, confirming the importance of specifically targeting the L-IFG in our \textit{in silico} experiments.  Additionally, consistent with previous findings \cite{Muldoon2016}, we observed that the patterning of activation as a result of stimulation of randomly selected brain regions varied with the selection of brain regions and across subjects. These findings support the possibility that in the future our modeling approach can be used to help design and optimize therapeutic strategies that use external activation such as TMS to treat neurological disorders \cite{Eldaief2013}.

\subsection*{Predicting task performance post-TMS}
\begin{figure}
\centering
\includegraphics[width=0.8\linewidth]{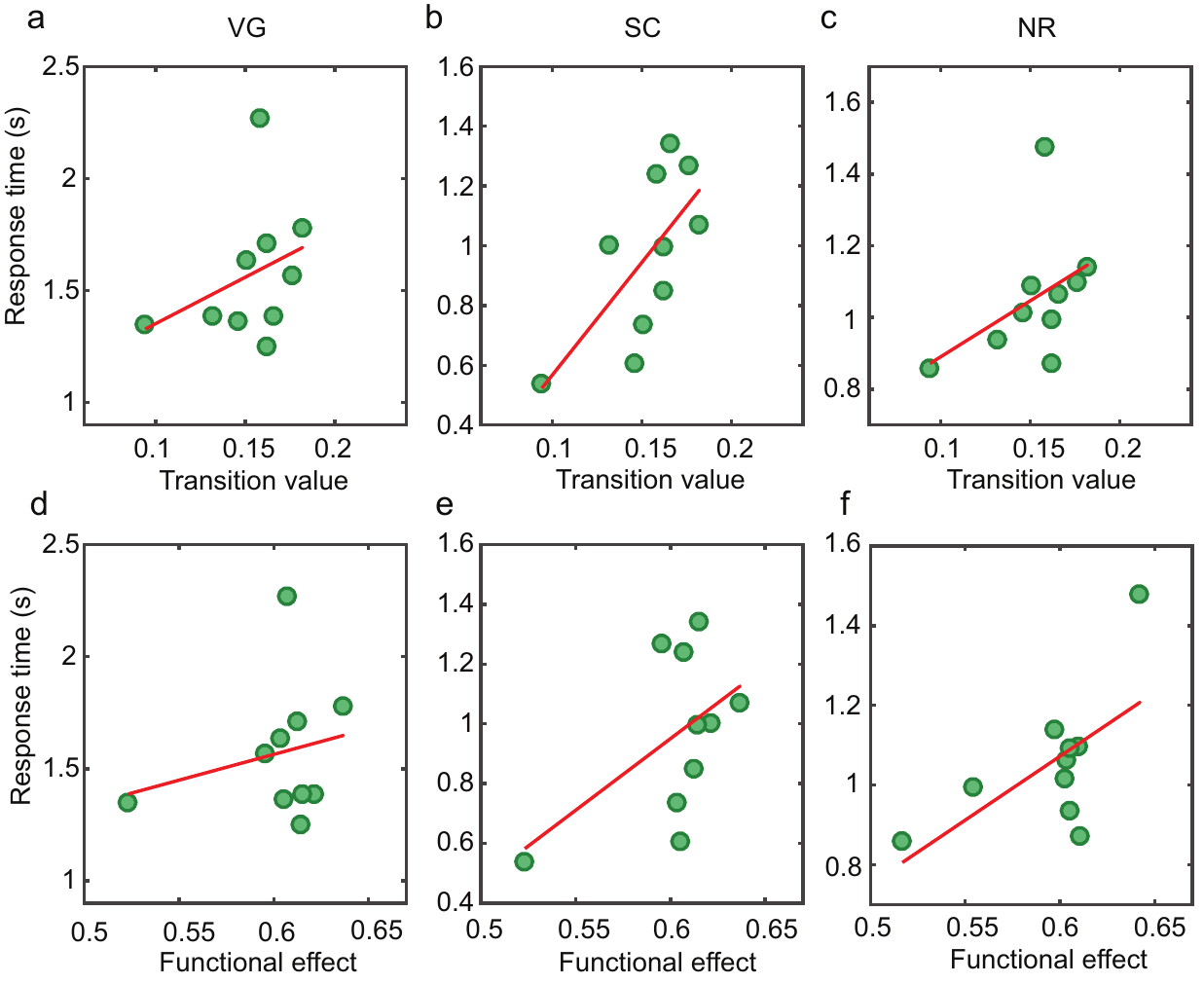}
\caption{{\bf Correlations between model features and task performance post-TMS.} 
(a-c) Task performance post-TMS versus the model transition value for three tasks: (a) verb generation (VG), (b) sentence completion (SC), and (c) number reading (NR). (d)-(f) Task performance post-TMS versus the functional effect within task-specific circuits for three tasks: (a) verb generation (VG), (b) sentence completion (SC), and (c) number reading (NR). The red lines represent a linear fit to the data for visual guidance. The corresponding Pearson correlation coefficients are given in Table \ref{table:corr_post_tms}. 
There is a significant correlation between the global transition value and task performance for the SC task ($r=0.68$, $p=0.03$) and between the functional effect and task performance of the NR task ($r=0.63$, $p=0.05$).}
\label{fig:tms}
\end{figure}

We also asked if our computational model could predict individual performance after the application of experimentally applied TMS targeted at the L-IFG. The underlying mechanisms of how TMS affects the brain are not well understood, but it is believed that TMS locally influences neuronal firing which can then propagate within the brain through inter-regional neuroanatomical pathways \cite{{Gu2015},{Muldoon2016}}. We therefore examined the correlation between model features and behavioral performance during the post-TMS task. As seen in Fig. \ref{fig:tms} and Table \ref{table:corr_post_tms}, while we still observe statistically significant positive correlations between the global coupling parameter and performance in the SC task ($r =0.68$, $p<0.03$) and between the functional effect within the task circuit and performance in the NR task ($r=0.63$, $p=0.05$), in both cases, the strength of the correlation is decreased when compared to correlations with task performance before TMS. Speculatively, the fact that the correlation strength is weakened post-TMS potentially reflects contributions of noise to the system mediated through inhibitory stimulation to the L-IFG.

\begin{table}[!ht]
\centering
\caption{
{\bf Correlations between post-TMS task performances and model features.}}
\begin{tabular}{|l|cc|cc|cc|}
\hline
&\multicolumn{2}{c|}{VG}  &  \multicolumn{2}{c|}{SC} &  \multicolumn{2}{c|}{NR}\\ 
 Model feature& r & p & r & p& r & p  \\ \hline
Transition value& 0.35 & 0.33& \textbf{0.68} &  \textbf{0.03} & 0.45& 0.20\\ \hline
Functional effect (global brain) & 0.19&0.59&0.39 &0.27 &0.08 &  0.82\\ \hline
Functional effect (task circuit) &0.23 & 0.52 & 0.52 & 0.12 & \textbf{0.63}&  \textbf{0.05}\\ \hline
Functional effect (outside the task circuit) &0.18& 0.62 & 0.37 &0.29 & 0.04& 0.91\\ \hline
\end{tabular}
\begin{flushleft} The variables $r$ and $p$ denote the Pearson correlation coefficient and associated \textit{p}-value, respectively. VG = verb generation, SC = sentence completion, and NR = number reading. 
\end{flushleft}
\label{table:corr_post_tms}
\end{table}

\subsection*{Correlations with graph theoretical measures of network connectivity}
Finally, to be able to draw a direct connection between anatomical brain connectivity and behavior, we asked if the predictions made by our computational model of brain dynamics could be revealed using only graph theoretical measures of network structure applied directly to the anatomical connectivity matrices alone (in the absence of model dynamics). We therefore calculated three measures of network structure that have been shown to be related to the spread of synchronization in networks: the average degree, inverse spectral radius, and synchronizability (see Materials and Methods). The average degree measures the average strength of connections per brain region, while the spectral radius and synchronizability are related to the eigenspectrum of the adjacency matrix or graph Laplacian respectively, and relate to the ease and extent to which the network is expected to synchronize under the assumption that regions behaved as oscillators.

In Table \ref{table:network_features}, we report the Pearson correlation coefficients for these measures of network structure with task performance both before and after TMS was applied. Both the average degree and the inverse spectral radius show significant correlations with task performance for the SC task, and as observed earlier, show a decrease in their correlation magnitude after the application of TMS. Both of these measures are known to relate to the ease of global synchronization in a network \cite{{Restrepo2005},{Meghanathan2014}}, and therefore this finding is somewhat expected given our prior observations about $c_5^{T}$. Interestingly, we find that the synchronizability of the network does not correlate with performance of the SC task. Instead, this measure, which has also been shown to be more sensitive to changes in local network structure \cite{{Huang2008},{Chen2008}}, shows a correlation with performance of the NR task that increases after the application of TMS. However, this correlation is weaker than that observed with the local measurements of the functional effect within the task circuit reported earlier in Table \ref{table:corr_pre_tms}.

\begin{table}[!ht]
\centering
\caption{
{\bf Correlations between task performances and structural network features.}}
\begin{tabular}{|l|cc|cc|cc|cc|cc|cc|}
\hline
& \multicolumn{4}{c|}{VG}  &  \multicolumn{4}{c|}{SC}  &  \multicolumn{4}{c|}{NR} \\
Network feature& \multicolumn{2}{c|}{Before-TMS}  &  \multicolumn{2}{c|}{Post-TMS}  & \multicolumn{2}{c|}{Before-TMS}  &  \multicolumn{2}{c|}{Post-TMS} & \multicolumn{2}{c|}{Before-TMS}  &  \multicolumn{2}{c|}{Post-TMS} \\
 &  r & p & r & p &  r & p &  r & p &   r & p &   r & p  \\
\hline
Average degree		& -0.25&0.48&-0.14&0.70&	\textbf{-0.84}&0.002&\textbf{-0.65}&	0.04&		-0.10&0.78&	-0.26&0.47\\
Inverse spectral radius 		& 0.29&0.42&0.36&0.31&		\textbf{0.87}&0.001&\textbf{0.70}&0.02&		0.31&	0.38&	0.48&	0.16 \\
Synchronizability			&-0.02&0.96&0.51&0.13&		0.58&	0.08&\textbf{0.64}&0.05&					\textbf{0.68}&0.03&\textbf{0.73}&0.02\\
\hline
\end{tabular}
\begin{flushleft} The variables $r$ and $p$ denote the Pearson correlation coefficient and associated \textit{p}-value, respectively. VG = verb generation, SC = sentence completion, and NR = number reading. 
\end{flushleft}
\label{table:network_features}
\end{table}

\section*{Discussion}

We presented a computational model of individualized brain dynamics that allows us to make predictions about cognitive task performance based on the variability of anatomical brain connectivity between people. By analyzing global model excitability and localized patterns in the spread of a targeted stimulation in our data-driven model, we explain individual variability in the performance of cognitively demanding tasks and establish the significant role of brain network organization in driving individual behavior. 

We examined three cognitive tasks and found that different measures of model dynamics gave rise to correlations with task performance.  Our results indicate that as the complexity of the cognitive task increases, a larger portion of the brain network, including interhemispheric connections, contribute in determining the overall response.  The SC task, which has additional language processing demands compared to the other two tasks, requires understanding different words, constructing the overall meaning of the sentence, and determining the appropriate word to fit in the sentence. For this complex task, a global parameter of the model, $c_5^{T}$, which could be related to the overall excitability of the network, predicted individual task performance. We also observed significant correlations between global network properties of the anatomical connectivity matrices (specifically the average degree and inverse spectral radius) and performance of the SC task. The fact that these global measures are correlated with performance in the SC task but not the other two tasks possibly reflects the global nature and complexity of the network of brain regions required to complete a sentence \cite{Just1996}.  

On the other hand, performances on the NR task, which has been established to be more localized in terms of the brain regions involved \cite{Roux2008}, can be predicted by tracking the spread of a targeted stimulation within a task-specific circuit in the brain. Further, we observed that a higher functional effect of the stimulation correlates with a higher response time (poorer performance) on the NR task.  In our model, a high functional effect indicates a high degree of synchronization within the task circuit. This heightened level of synchronization could effectively constrain the degrees of freedom for brain circuit activity and computation, resulting in poorer task performance.  Better task performance might be achieved through an integration and segregation mechanism which allows brain regions to mutually communicate but does not require synchrony over a longer time scale. Such a modular functional organization has been associated with efficient learning of motor skills \cite{Bassett2011b}.

Interestingly, no tested measures of model dynamics or structural network features were able to explain the performance of the VG task.  It is likely that VG requires a different coupling mechanism of global and local dynamics which is not captured in the model features constructed here. For example, while verb generation and sentence completion are both ``open ended'' language tasks, SC requires our ability to accumulate the probability of a response over an entire sentence, whereas VG is cued by a single word.  Thus, even if both involve the L-IFG, SC might recruit quite a few more distributed resources, while VG might require more focal resources.  It is possible that our selection of regions comprising the language task circuit was not sensitive enough to these focal areas.  We propose that this problem could potentially be resolved by a more extensive research effort that combines experimental and theoretical approaches from cognitive and network neuroscience in order to better understand the specific circuitry involved in task completion.  

A strength of our modeling approach is that the construction of localized task circuits is not limited to those involved in the tasks studied here, and could easily be extended to represent different brain circuitry. Ultimately, our model magnifies the effects of differences in the observed anatomical brain structure, and it is not always feasible to measure differences in network structure in terms of general network statistics.  While some network statistics also provided predictive power of task performance, the strength of the correlation was only comparable for measures of global network structure, and our computational model can generate measures of structure and dynamics specific to the task that is being performed.

It should also be noted that the functional measures examined here (for both global and task circuit specific calculations) involved measuring the spread of synchronization after computational stimulation of the L-IFG.  The choice to stimulate the L-IFG represents essential \emph{a priori} knowledge from cognitive neuroscience studies, and stimulating randomly chosen sets of brain regions within the task circuits did not produce significant correlations with task performances. The ability to integrate knowledge from cognitive neuroscience into computational experiments represents a strength of our data-driven modeling approach, as it makes the model flexible and opens the door to studies across a range of cognitive paradigms.

Although the current study was constrained by its small subject population size, it is encouraging that despite this limitation, we were still able to make certain predictions about task performance.  These findings should therefore promote the use of similar data-driven modeling approaches in larger data sets with more subjects and a higher diversity of task conditions.  The use of personalized brain network models will serve as an increasingly valuable tool to establish explicit links between brain connectivity, dynamics, and behavior, and to develop personalized therapeutic strategies.

\section*{Materials and Methods}

\subsection*{Subjects and cognitive tasks}
Ten healthy individuals (mean age = 25.4, St.D. = 4.5, 6 female) from a larger neuroimaging study \cite{Betzel_2016} returned to participate in the present study. All procedures were approved in a convened review by the University of Pennsylvania's Institutional Review Board and were carried out in accordance with the guidelines of the Institutional Review Board/Human Subjects Committee, University of Pennsylvania. All participants volunteered with informed consent in writing prior to data collection.

Participants performed two open-ended language tasks and one closed-ended number naming task. The language tasks included a verb generation task \cite{KRIEGERREDWOOD201424} and a sentence completion task \cite{Snyder2008}. For the verb generation task, subjects were instructed to generate the first verb that came to mind when presented with a noun stimulus (e.g., `cat'). The verb could be either something that the noun does or something that can be done with the noun. Response times (RTs) were collected from the onset of the noun cue to the onset of the verb response. For the sentence completion task, subjects were presented with a sentence, for example ``They left the dirty dishes in the -----?'', and were instructed to generate a single word that appropriately completes the sentence, such as 'sink'. Words in the sentences were presented serially in 1 s segments consisting of one or two words. RTs were computed as the latency between the onset of the last segment, which always contained two words (i.e., a word and an underline), and the onset of the participant's response. For all items in the sentence completion task, items in the high vs. low selection demand conditions were matched on retrieval demands (association strength) \cite{Snyder2008}. For both language tasks, each trial began with the presentation of a fixation point (+) for 500 ms, followed by the presentation of the target stimulus, which remained on the screen for 10 s until the subject made a response. Subjects were given an example and five practice trials in the first administration of each language task (i.e., before TMS), and were reminded of the instructions before performing the task a second time (i.e., after TMS). In each of the before and after TMS conditions, subjects completed 50 trials for a total of 100 trials.

In the number reading task, participants produced the English names for strings of Arabic numerals presented on the screen. On each trial, a randomized number (from tens of thousands to millions; e.g., 56395, 614592, 7246856) was presented in black text on a white background. The numbers were uniformly distributed over three lengths (17 per length for each task administration). The position of items on the screen was randomized between the center, left, and right of the screen to reduce the availability of visual cues to number length and syntax \cite{KRIEGERREDWOOD201424}. RTs were collected from the onset of the stimulus presentations to the onset of the subject's response. The number appeared in gray following the detection of a response (i.e., voice key trigger), and remained on the screen thereafter to reduce the working memory demands required for remembering the digit string. At the start of the experiment, subjects performed 50 trials of the number naming task to account for initial learning effects \cite{KRIEGERREDWOOD201424}. Prior to performing the task for the first time, subjects were given an example and five practice trials, and were later reminded of the instructions before performing the task a second (i.e., before TMS) and a third (i.e., after TMS) time. In each of the before and after TMS conditions, subjects completed 51 trials for a total of 102 experimental trials. The items for the verb generation task were identical to those used in \cite{Snyder2011} and the items for the sentence completion task were those from \cite{Snyder2014}. The difficulty of items was sampled to cover a distribution of values computed via latent semantic analysis (LSA) applied to corpus data. 

Verbal responses for all tasks were collected from a computer headset microphone. The microphone was calibrated to reduce sensitivity to environment background noise prior to the collection of data for each session such that the recording software was not triggered without clear verbalizations. List order (before or after TMS) was counterbalanced across participants. Item presentation order within each task was fully randomized across participants.  Task performance was assessed based on the subject's median response time across all the trials.

\subsection*{Transcranial Magnetic Stimulation}
The Brainsight system (Rogue Research, Montreal) was used to co-register MRI data with the location of the subject and the TMS coil. The stimulation site was defined as the posterior extent of the pars triangularis in each individual subject's registered T1 image. A Magstim Super Rapid2 Plus1 stimulator (Magstim; Whitland, UK) was used to deliver continuous theta burst stimulation (cTBS) via a 70 mm diameter figure-eight coil. cTBS was delivered at 80\% of each participant's active motor threshold \cite{Huang2005}. Each participant's threshold was determined prior to the start of the experimental session using a standard up-down staircase procedure with stimulation to the motor cortex [M1].

\subsection*{Human DSI data acquisition and preprocessing}
Diffusion spectrum images (DSI) were acquired for a total of 10 subjects along with a T1-weighted anatomical scan at each scanning session, in line with previous work \cite{Gu2015}. DSI scans sampled 257 directions using a Q5 half-shell acquisition scheme with a maximum b-value of 5,000 and an isotropic voxel size of 2.4 mm. We utilized an axial acquisition with the following parameters: repetition time (TR) = 5 s, echo time (TE)= 138 ms, 52 slices, field of view (FoV) (231, 231, 125 mm). DSI data were reconstructed in DTI Studio (www.dsi-studio.labsolver.org) using q-space diffeomorphic reconstruction (QSDR)\cite{YEH20111054}. QSDR first reconstructs diffusion-weighted images in native space and computes the quantitative anisotropy (QA) in each voxel. These QA values are used to warp the brain to a template QA volume in Montreal Neurological Institute (MNI) space using the statistical parametric mapping (SPM) nonlinear registration  algorithm. Once in MNI space, spin density functions were again reconstructed with a mean diffusion distance of 1.25 mm using three fiber orientations per voxel. Fiber tracking was performed in DSI Studio with an angular cutoff of $35^{\circ}$, step size of 1.0 mm, minimum length of 10 mm, spin density function smoothing of 0.0, maximum length of 400 mm and a QA threshold determined by DWI signal in the colony-stimulating factor. Deterministic fiber tracking using a modified FACT algorithm was performed until 1,000,000 streamlines were reconstructed for each individual.

Anatomical (T1) scans were segmented using FreeSurfer \cite{FISCHL2012774} and parcellated using the connectome mapping toolkit \cite{CAMMOUN2012386}. A parcellation scheme including n=234 regions was registered to the B0 volume from each subject's DSI data. The B0 to MNI voxel mapping produced via QSDR was used to map region labels from native space to MNI coordinates. To extend region labels through the grey-white matter interface, the atlas was dilated by 4 mm \cite{Cieslak2014}. Dilation was accomplished by filling non-labelled voxels with the statistical mode of their neighbors' labels. In the event of a tie, one of the modes was arbitrarily selected. Each streamline was labeled according to its terminal region pair.

\subsection*{Construction of anatomical brain networks}
To construct the subject-specific anatomical connectivity networks used as a basis for our computational model, we segmented the brain into 234 regions (network nodes) based on the Lausanne atlas \cite{{CAMMOUN2012386},{Hagmann2008}}.  As in prior studies, we define pairwise connection weights between nodes based on the number of streamlines connecting brain regions and normalized by the sum of the volumes of the nodes \cite{{Muldoon2016}, {Hagmann2008}}.  This procedure results in a sparse, weighted, undirected structural brain network for each subject ($N=10$), where network connections represent the density of white matter tracts between brain regions (Fig. \ref{fig:schematic}b).

\subsection*{Computational model of brain dynamics}
In our data-driven network model, regional brain dynamics are given by Wilson-Cowan oscillators \cite{{Wilson1972},{Muldoon2016}}. In this biologically motivated model of neuronal populations, the fraction of excitatory and inhibitory neurons active at time $t$ in the $i^{th}$ brain region are denoted by $E_i(t)$ and $I_i(t)$ respectively, and their temporal dynamics are given by:  
\begin{equation}
\tau\frac{dE_i}{dt}=-E_i(t)+(S_{E_m} - E_i(t))S_E\Big(c_1E_i(t)-c_2I_i(t) \\+ c_5\sum\limits_{j}A_{ij}E_j(t-\tau_d^{ij})+P_i(t)\Big)+\sigma w_i(t),
\end{equation}

\begin{equation}
\tau\frac{dI_i}{dt}=-I_i(t)+(S_{I_m} - I_i(t))S_I\Big(c_3E_i(t)-c_4I_i(t) \\+ c_6\sum\limits_{j}A_{ij}I_j(t-\tau_d^{ij})\Big)+\sigma v_i(t),
\end{equation}

where

\begin{equation}
S_{E,I}(x) = \frac{1}{1+e^{(-a_{E,I}(x-\theta_{E,I})}} - \frac{1}{1+e^{a_{E,I}\theta_{E,I}}}.
\end{equation}

\noindent We note that $A_{ij}$ is an element of the subject-specific coupling matrix, $A$, whose value is the connection strength between brain regions $i$ and $j$ as determined from DSI data (see above). The global strength of coupling between brain regions is tuned by excitatory and inhibitory coupling parameters $c_5$ and $c_6$ respectively. We fix $c_6=c_5/4$, representing the approximate ratio of excitatory to inhibitory coupling. $P_i(t)$ represents the external inputs to excitatory state activity and is used to perform computational stimulation experiments (see below).  The parameter $\tau_d^{ij}$ represents the communication delay between regions $i$ and $j$. If the spatial distance between regions $i$ and $j$ is $d_{ij}$, $\tau_d^{ij}=d_{ij}/t_d$, where $t_d = 10m/s$ is the signal transmission velocity. Additive noise is input to the system through the parameters $w_i(t)$ and $v_i(t)$ which are derived from a normal distribution and $\sigma = 10^{-5}$. Other constants in the model are biologically derived: $c_1 = 16$, $c_2 = 12$, $c_3 = 15$, $c_4 = 3$, $a_E = 1.3$, $a_I = 2$, $\theta_E = 4$, $\theta_I = 3.7$, $\tau = 8$ as described in references \cite{{Wilson1972},{Muldoon2016}}. 
 
To numerically simulate the dynamics of the system we use a second order Runge Kutta method with step size $0.1$ and initial conditions $E_i(0)=I_i(0)=0.1$.  All analysis is performed after allowing the system to stabilize for $1$s.

\subsection*{Assessing individual variability}
In order to calculate the model transition values for each individual, we ran $1$s simulations (after allowing the system to stabilize) for a range of $c_5$ parameters ($0.05\leq c_5\geq 0.25$, with a step-size of $0.001$) in which no external input was applied ($P = 0$). The average excitatory activity was recorded  for each region as a function of $c_5$ as shown in Fig. \ref{fig:schematic}c. The value of $c_5$ at which we observed a sudden increase in the average activity (marked by an arrow in Fig. \ref{fig:schematic}c), was identified as the transition value, $c_5^T$. 

Structural variability was measured by calculating the standard deviation of a given connection strength $A_{ij}$  between brain regions $i$ and $j$ for all the subjects and then normalizing by the average connection strength, $<A_{ij}>$.

\subsection*{Targeted computational activation} 
To activate (stimulate) a particular brain region, we applied a constant external input $P_i = 1.15$ which drives the regional activity into a limit cycle as shown in Figure \ref{fig:functional}b. Before targeted activation, the global coupling parameter $c_5$ was fixed just below its transition value $c_5^T$ (which differs between subjects) to perturb the system in a state with maximum sensitivity for perturbations. When stimulating the L-IFG, we simultaneously activated four brain regions (\textit{pars orbitalis} (POB), \textit{pars triangularis} (PTA) and \textit{pars opercularis} (POC, two regions), Fig. \ref{fig:functional}a) by applying an external input of $P=1.15$. 

\subsection*{Functional connectivity and functional effect of stimulation}
To quantify the spread of computational stimulation, functional connectivity was determined by calculating the pairwise maximum normalized cross-correlation \cite{{Kramer2009},{Feldt2007}} between the excitatory activity $E_i(t)$ and $E_j(t)$, for brain regions $i$ and $j$. We used a window size of $1$s and calculated correlations over a maximum lag of $250$ms. 

In order to quantify the functional effect of stimulation, we considered the regional dynamics within a window of $2$s once the system is stabilized after the initial transient period. The system was first allowed to evolve without any external input ($P_i = 0$) for $1$s and then an external input of $P_i=1.15$ was applied for $1$s to the set of brain regions selected for activation or stimulation. Functional connectivity was calculated separately for the stimulation-free period and the period of stimulation.  We then calculated the difference between the pairwise values of functional connectivity measured during and before stimulation, resulting in a matrix that describes the pairwise changes in functional connectivity resulting from the stimulation.  The functional effect of stimulation \cite{Muldoon2016} was then calculated using this matrix either globally by averaging over the entire matrix, or within a task circuit by averaging within the task circuit, or outside of a task circuit by averaging over regions outside of the task circuit.

\subsection*{Defining task circuits}
To construct a task dependent localized measure, we followed the work of Roux et al. \cite{Roux2008} to identify the possible brain regions involved in reading alphabets and numbers. We extended their findings to propose that the brain regions involved in reading alphabets contribute towards performing VG and SC tasks, and the brain regions involved in reading numbers contribute towards performing the NR task. We mapped these regions to the Lausanne atlas and constructed possible task circuits involved in VG and SC (alphabets-related, Fig. \ref{fig:circuits}a), and NR (number-related, Fig. \ref{fig:circuits}b). These circuits are also consistent with other studies mapping brain regions involved in language processing \cite{{Dronkers2004},{Turken2011},{NathanielJames2002},{Holland2001}}.

\subsection*{Randomizing network structure and stimulation effect} 

To assess the effect of global network organization, we repeated our entire computational analysis with randomized anatomical connectivity matrices. Anatomical connectivity matrices were randomized separately for each individual in order to alter the original brain network organization. Here, we preserved the overall connectivity distribution but randomly reassigned connection strength values to each pair of the brain regions from this distribution.

To test the specificity of our results to the choice of regions comprising the task circuit, we constructed $10,000$ sub-networks by randomly picking the same number of brain regions as in the NR circuit ($22$) from all of the $234$ brain regions. We then stimulated the L-IFG and calculated the functional effect within these randomly constructed sub-networks for each individual. These sub-networks had varying degree of overlap with the actual task circuit, ranging from $0$ to $35\%$. We observed that only $1.8\%$ of the random sub-networks produced a significant correlation between the localized functional effect and task performance with $r > 0.5$ and $p < 0.05$, indicating that our results are indeed related to the specific construction of the task circuit. 

In order to assess the importance of stimulating the L-IFG for the cognitive tasks chosen in this particular work, we compared our findings with those when stimulation was applied to a group of regions randomly chosen (under spatial constraints) within the actual task circuits with a regional brain volume equivalent to L-IFG ($4$ brain regions). The $4$ brain regions were chosen such that they formed a continuous spatial volume within the brain. We could identify $7$ such volumes within the task circuits that we constructed, and used these $7$ alternate stimulation sites to assess the specificity of our findings to the choice of stimulation site.  The model did not produced significant predictions when any of these $7$ volumes were stimulated, indicating that the model is indeed sensitive to the choice of stimulation site.

\subsection*{Network statistics} 
We calculated network statistics for each subject using the structural matrices derived from their DSI data. The weighted degree centrality, $k_i$, for a given region $i$ is defined as $k_i= \sum_{j=1}^{234}A_{ij}$ \cite{Newman_book}. The average across the degree centralities of all network nodes was used to obtain the average degree of a given subject. The spectral radius, $S_r$, is given by the maximum eigenvalue of the connectivity matrix $A$, $S_r=\lambda|_{Max}$,  and is a global measure of network structure that is related to the spread of synchronization in a network \cite{{Restrepo2005},{Meghanathan2014}}.  Synchronizability, $S$, is defined as, $S=\frac {\lambda_2^L}{\lambda_{Max}^L}$, where ${\lambda_2^L}$ and ${\lambda_{Max}^L}$ denote the second smallest and the largest eigenvalue of the Laplacian matrix $L$ ($L =  D - A$, where D is the degree matrix of $A$) \cite{{Huang2008},{Chen2008}}.

\section*{Acknowledgments}

This work is supported by the U.S. Army Research Laboratory through contract numbers W911NF-10-2-0022 and W911NF-16-2-0158 from the U.S. Army research office. DSB acknowledges MacArthur Foundation and the Alfred P. Sloan Foundation. JDM acknowledges funding from NIH and NIMH (award no. 1-DP5 OD021352-01). The content is solely the responsibility of the authors and does not necessarily represent the official views of the funding agencies.

%

\end{document}